\documentclass[journal=jacsat,manuscript=article]{achemso}

\usepackage{graphicx} 
\usepackage{subcaption} 

\usepackage{enumitem}
\setlist{noitemsep,leftmargin=*,topsep=0pt,parsep=0pt}

\usepackage{hyperref}
\hypersetup{
colorlinks=true,
urlcolor= blue,
citecolor=blue,
linkcolor= blue,
}

\title{Persistence of structural distortion and bulk band Rashba splitting in SnTe above its ferroelectric critical temperature}

\author{Frédéric Chassot}
\email{frederic.chassot@unifr.ch}
\affiliation{Department of Physics and Fribourg Center for Nanomaterials, Université de Fribourg, Fribourg, Switzerland}

\author{Aki Pulkkinen}
\affiliation{Department of Physics and Fribourg Center for Nanomaterials, Université de Fribourg, Fribourg, Switzerland}
\alsoaffiliation{New Technologies-Research Center, University of West Bohemia, Plzeň, Czech Republic}

\author{Geoffroy Kremer}
\affiliation{Department of Physics and Fribourg Center for Nanomaterials, Université de Fribourg, Fribourg, Switzerland}
\alsoaffiliation{Institut Jean Lamour, UMR 7198, CNRS-Université de Lorraine, Campus ARTEM, 2 allée André Guinier, BP 50840, 54011 Nancy, France}

\author{Tetiana Zakusylo}
\affiliation{Institut für Halbleiter-und Festkörperphysik, Johannes Kepler Universität, Linz, Austria}

\author{Gauthier Krizman}
\affiliation{Institut für Halbleiter-und Festkörperphysik, Johannes Kepler Universität, Linz, Austria}

\author{Mahdi Hajlaoui}
\affiliation{Institut für Halbleiter-und Festkörperphysik, Johannes Kepler Universität, Linz, Austria}

\author{J. Hugo Dil}
\affiliation{Institute of Physics, Ecole Polytechnique Fédérale de Lausanne, Lausanne, Switzerland}
\alsoaffiliation{Photon Science Division, Paul Scherrer Institut, Villigen, Switzerland}

\author{Juraj Krempask\'y}
\affiliation{Photon Science Division, Paul Scherrer Institut, Villigen, Switzerland}

\author{Ján Minár}
\affiliation{New Technologies-Research Center, University of West Bohemia, Plzeň, Czech Republic}

\author{Gunther Springholz}
\affiliation{Institut für Halbleiter-und Festkörperphysik, Johannes Kepler Universität, Linz, Austria}

\author{Claude Monney}
\email{claude.monney@unifr.ch}
\affiliation{Department of Physics and Fribourg Center for Nanomaterials, Université de Fribourg, Fribourg, Switzerland}

\begin{document}

\date{31.08.2023}

\begin{abstract}
The ferroelectric semiconductor $\alpha$-SnTe has been regarded as a topological crystalline insulator and the dispersion of its surface states has been intensively measured with angle-resolved photoemission spectroscopy (ARPES) over the last decade. However, much less attention has been given to the impact of the ferroelectric transition on its electronic structure, and in particular on its bulk states.
Here, we investigate the low-energy electronic structure of $\alpha$-SnTe with ARPES and follow the evolution of the bulk-state Rashba splitting as a function of temperature, across its ferroelectric critical temperature of about $T_c\sim 110$ K. Unexpectedly, we observe a persistent band splitting up to room temperature, which is consistent with an order-disorder contribution to the phase transition that requires the presence of fluctuating local dipoles above $T_c$. We conclude that no topological surface state can occur at the (111) surface of SnTe, at odds with recent literature.
\end{abstract}

\maketitle
\section{\label{sec:Intro}Introduction}
Semiconductors based spintronics materials are one of the most promising playground for modern applications and technologies \cite{wolf_spintronics_2001, book}. Among these materials, the class IV-VI semiconductors are  particularly interesting because they can combine semiconductor properties with ferroelectricity. This is due to a spontaneous distortion of the crystalline lattice structure that leads to  a macroscopic electric polarization of the material \cite{rabe_modern_2007, noel_non-volatile_2020,noauthor_century_2020,Bhalla_2021}. In addition, the concomitant inversion symmetry breaking induces a momentum-dependent energy splitting in the electronic band structure, i.e. the so-called Rashba effect, which means that these bands are not anymore spin degenerate \cite{bychkov1984properties,Rotenberg_Spin-Orbit_1999}. 

In this framework, the electronic band structure of $\alpha$-GeTe, a ferroelectric Rashba semiconductor with a critical temperature $T_c=670$ K \cite{pawley_diatomic_1966}, has been investigated in details. Different surface states, surface resonances and bulk states have been identified using angle-resolved photoemission spectroscopy (ARPES) \cite{krempasky_disentangling_2016,PhysRevB.94.201403} and those studies have led to the observation of one of the largest Rashba parameters \cite{di_sante_electric_2013}. Its potential for application is then particularly large, e.g. with the ability to enhance spin Hall conductivity \cite{wang_spin_2020}, to control the spin-to-charge conversion, to store information in a non-volatile way \cite{rinaldi_evidence_2016,picozzi_ferroelectric_2014,liebmann_giant_2016,krempasky_entanglement_2016,krempasky_operando_2018} or also to manipulate the crystal distortion and thus the ferroelectricity using intense femtosecond pulses \cite{kremer_field-induced_2022}. The isostructural compound SnTe has similar properties than GeTe and it shows ferroelectricity typically below 100 K \cite{iizumi_phase_1975}. 

For both GeTe and SnTe, the nature of the ferroelectric transition is still subject to debate, even though it has attracted a lot of attention in the literature. Early studies using neutron diffraction or Raman scattering to reveal the atomic structure and related phonons \cite{iizumi_phase_1975,pawley_diatomic_1966,brillson_raman_1974} suggested that the transition temperature could strongly depend on the number of Sn (Ge) vacancies and that the transition is of second order\cite{kobayashi_carrier-concentration-dependent_1976,mazelsky_phase_1962,pawley_diatomic_1966}. Subsequently, this was confirmed by theoretical  \cite{littlewood_crystal_1980,rabe_ab_1985,salje_tin_2010,li_anomalous_2022} and experimental studies that demonstrated a phonon softening at $T_c$, indicating a displacive phase transition \cite{oneill_inelastic_2017}. 
However, extended x-ray absorption fine structure, x-ray scattering measurements and analysis of the pair distribution function, evidenced the persistence of a local rhombohedral lattice distortion above $T_c$, indicating the presence of local ferroelectric dipoles \cite{fornasini_local_2021,mitrofanov_local_2014,fons_phase_2010,matsunaga_order-disorder_2011}. However, this conclusion was criticized in another work based on the analysis of pair distribution function \cite{chatterji_anomalous_2015}, emphasizing that a vivid debate on the ferroelectric phase transition in SnTe still remains.

SnTe has been regarded as an outstanding representative of the class of topological crystalline insulators. However, the ferroelectric transition has also considerable effect on its topological properties. Symmetry arguments have been used to claim that in the paraelectric phase this semiconductor has gapless protected surface states \cite{fu_topological_2011}. This was first predicted theoretically \cite{hsieh_topological_2012,shi_111_2014} and the existence of linear-dispersive bands attributed to topologically protected surface states was later on confirmed by ARPES for the $(100)$ and $(111)$ surface-plane orientations\cite{tanaka_experimental_2012,tanaka_two_2013,yan_experimental_2014, zhang_arpes_2017, polley_observation_2016, PhysRevB.104.195403}. However, as shown by Pleakhanov \textit{et al.} \cite{plekhanov_engineering_2014}, the topological surface state does not subsist in a ferroelectric state on the (111) surface.

In the present work, we study the low-energy electronic structure of SnTe(111) across its ferroelectric phase transition with ARPES. Taking advantage of our high energy and momentum resolution and also of the unprecedented crystalline quality of our thin films, we reveal multiple states in the first eV below the Fermi level that have not been resolved in the literature so far. Based on one-step model photoemission calculations and photon-energy dependent ARPES measurements, we classify them as surface or bulk states. Most importantly, we systematically characterize the change of the Rashba splitting of bulk states as a function of temperature. We observe clear inconsistencies of the ferroelectric phase transition with a simple mean-field-like transition that can be explained with an order-disorder type contribution to the transition. Finally, we comment on its consequence for the topological properties of the (111) surface of SnTe.

\section{\label{sec:Results}Results}

\begin{figure}[H]
  \begin{subfigure}[b]{0.28\linewidth}
      \caption{Paraelectric}\label{fig:1a}
      \includegraphics[width=\columnwidth]{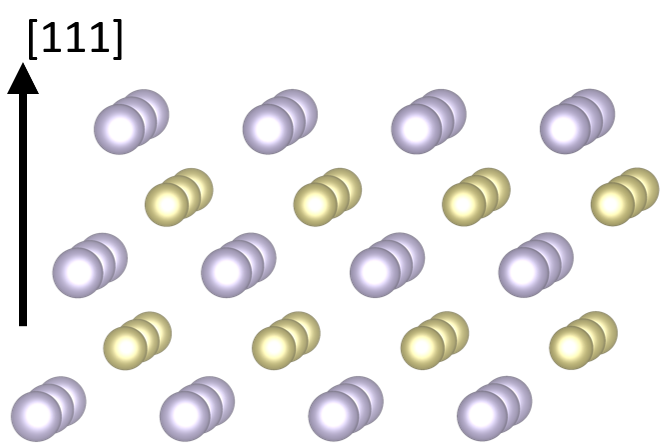}
   \end{subfigure}
   \begin{subfigure}[b]{0.28\columnwidth}
        \caption{Ferroelectric}\label{fig:1b}
        \includegraphics[width=\columnwidth]{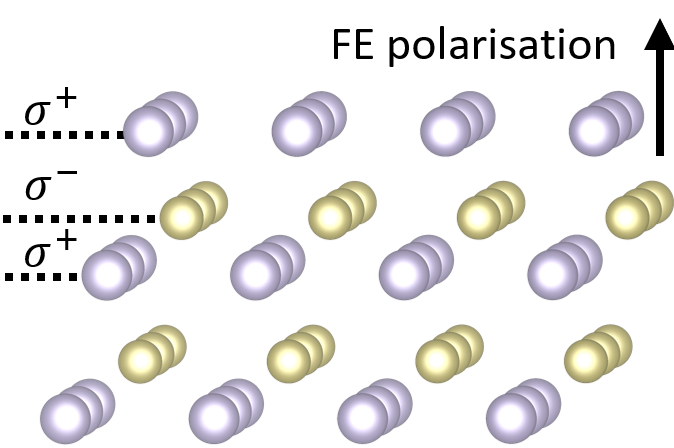}
    \end{subfigure}
\par\smallskip 
  \begin{subfigure}[b]{0.28\columnwidth}
  \caption{Brillouin zone}\label{fig:1c}
  \includegraphics[width=0.955\linewidth]{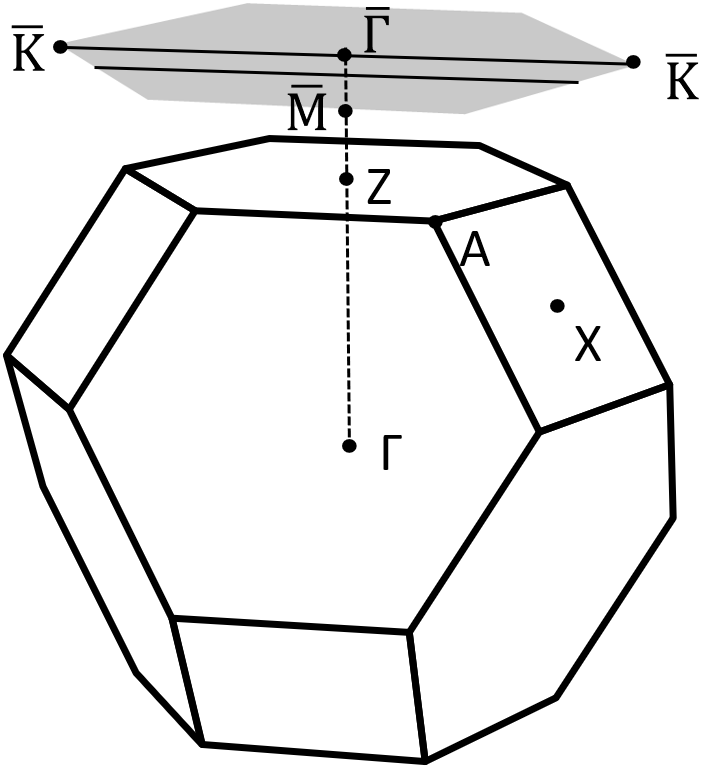} 
\end{subfigure}
\begin{subfigure}[b]{0.28\columnwidth}
  \caption{Bulk DFT} \label{fig:1d}
  \includegraphics[width=\linewidth]{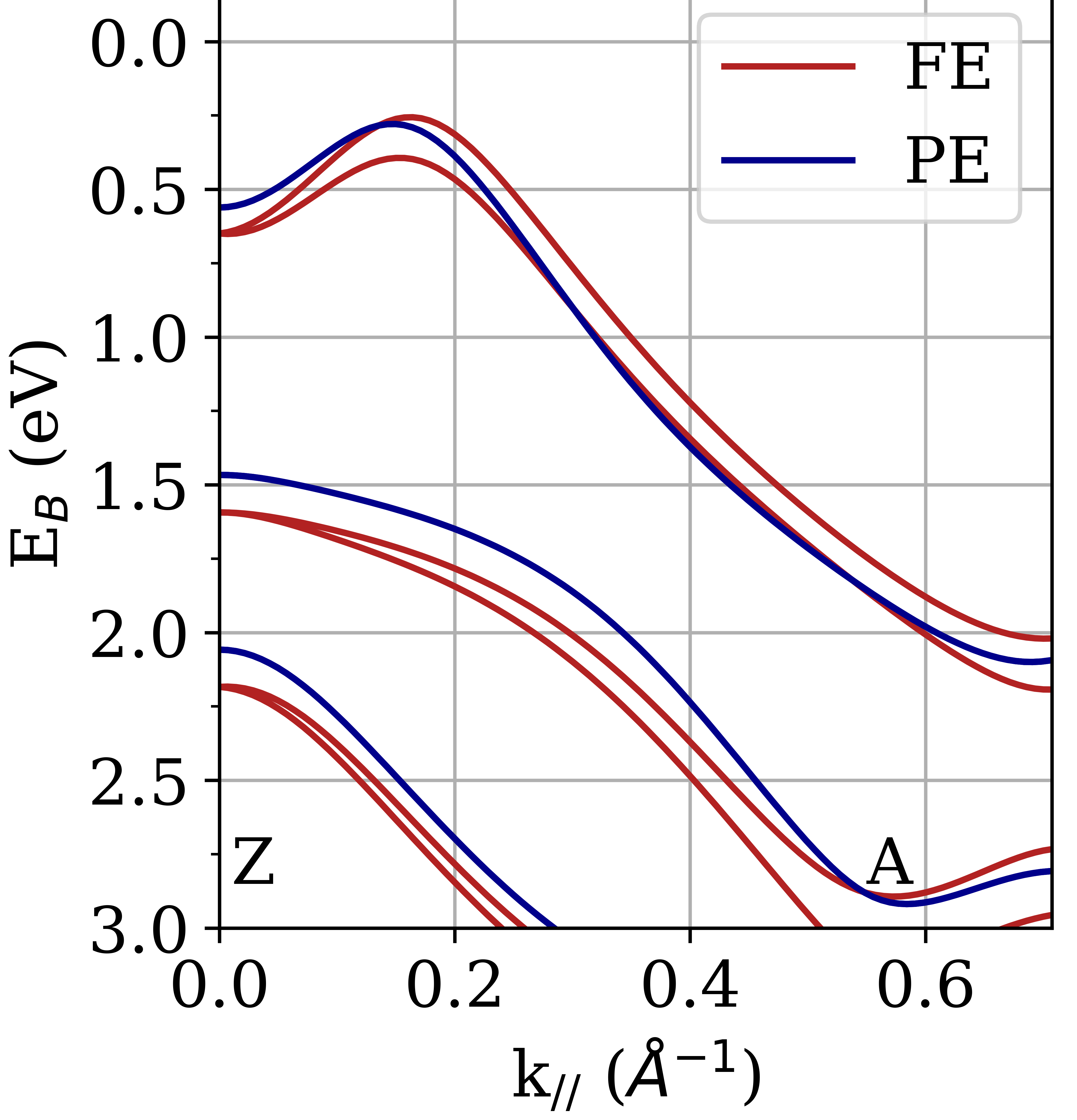}
\end{subfigure}
\caption{(a) Rocksalt structure of SnTe in the paraelectric state along the [111] crystalline direction. Blue (yellow) dots represent Sn (Te). (b) Rhombohedral structure of SnTe in the ferroelectric state (not to scale). Structures were generated with VESTA \cite{momma_vesta_2011}. (c) Bulk Brillouin zone of SnTe and its surface projected plane along the (111) direction. The black lines in the surface projected plane represent the directions of data presented in this work. (d) Bulk DFT band structure between the high symmetry points Z and A. A $k_{\perp}$ value of 50\% of the $\Gamma Z$ distance has been selected to approximate the reciprocal space plane sampled at a photon energy of 11.2 eV. Bands are plotted for the paraelectric (PE) cubic rocksalt (blue) and ferroelectric (FE) rhombohedral (red) structures.}
\label{fig:1}
\end{figure}

SnTe undergoes a transition from a paraelectric state, with a cubic rocksalt structure with equidistant stackings of Sn and Te layers along the [111] direction (space group $Fm\bar{3}m$, see Fig. \ref{fig:1a}), to a ferroelectric state with a rhombohedral structure (space group $R3m$, see Fig. \ref{fig:1b}) at low temperature around 100 K \cite{iizumi_phase_1975}. In the ferroelectric state, the bulk inversion symmetry is broken by a displacement of the Sn and Te lattice planes against each other, which leads to a non-zero electric dipole between the ionic charges $\sigma^+$ and $\sigma^-$ of the Sn and Te atoms.
This induces a Rashba-like splitting in the electronic structure, as can be seen by comparing the bulk DFT band structure in Fig. \ref{fig:1d} calculated for the paraelectric (blue bands) and the ferroelectric (red bands) states.
Our objective is to experimentally resolve this splitting and to follow its evolution as a function of temperature in order to characterize the ARPES signatures of the paraelectric-to-ferroelectric phase transition. 

\begin{figure}[H]
  \begin{subfigure}[b]{1\linewidth}
        \phantomcaption
       \label{fig:2a}
       \begin{center}
       \includegraphics[height=280pt]{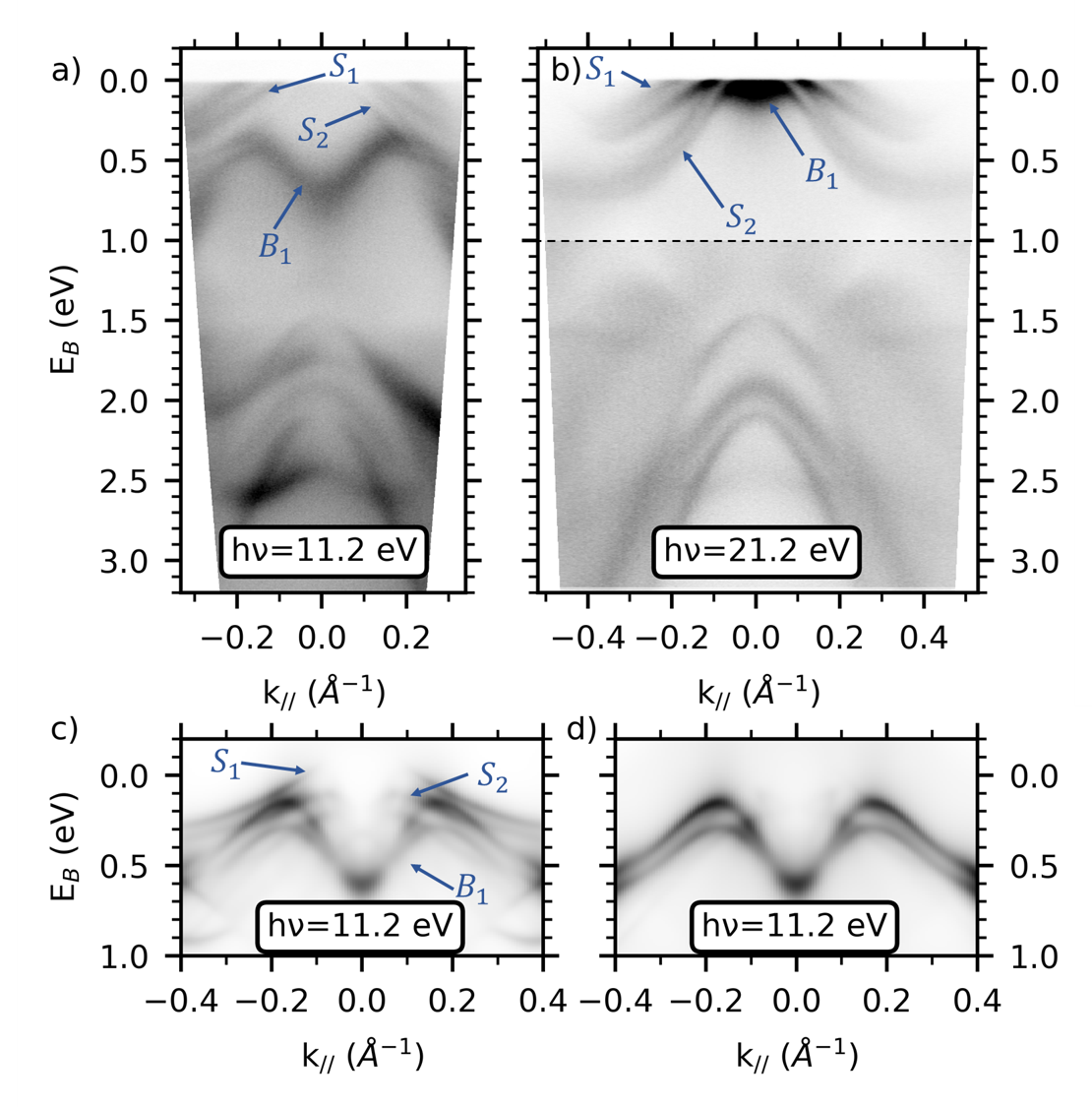}
       \end{center}
      \phantomcaption
       \label{fig:2b}
        \phantomcaption
       \label{fig:2c}
      \phantomcaption
       \label{fig:2d}
   \end{subfigure}
\caption{ARPES measurements along the $\overline{K\Gamma K}$ high-symmetry line at 30 K with a photon energy of (a) 11.2 eV and (b) 21.12 eV. The dashed line in panel (b) represents a change of saturation by a factor of 2. (c) One-step photoemission calculation for a slab geometry in a ferroelectric structure, with a photon energy of 11.2 eV and a Te termination. (d) Same calculation with a transparent barrier.} 
\label{fig:2}
\end{figure}

We have performed ARPES measurements of SnTe(111) along the $\overline{K\Gamma K}$ high-symmetry direction, which corresponds to the projection of the $AZA$ direction on the $(111)$ surface. Photoemission intensity maps obtained at 30 K with two different photon energies $h\nu$ are shown in Fig. \ref{fig:2a} for $h\nu=11.2$ eV and Fig. \ref{fig:2b} for $h\nu=21.2$ eV respectively. 
Thanks to the high energy resolution of our experiment and the high quality of our thin films, we distinguish several bands in the low-energy region, where previously only a linear dispersive band was resolved and attributed to a Dirac cone \cite{tanaka_two_2013,zhang_arpes_2017,yan_experimental_2014}. 
Near the Fermi level, we identify one surface state (labelled $S_1$) that appears for both photon energies at higher parallel momenta. At lower momenta, we observe another surface state ($S_2$) which partially overlaps with a bulk state ($B_1$) that disperses with photon energy. A more detailed series of photon-energy dependent ARPES measurements using synchrotron radiation is shown in the Supporting Information and corroborates these observations with Fig. \ref{fig:hvdep}.
Similar to the isostructural compound $\alpha$-GeTe \cite{krempasky_disentangling_2016}, we attribute the state $S_2$ to a surface resonance state.

To support our interpretation of the origin of the bands, we have performed DFT calculations using a semi-infinite slab geometry for the ferroelectric structure. The surface with a Te-termination and short bonds between the first Te and Sn planes gives the best agreement with the experimental data (see Supporting Information with Fig. \ref{fig:6}). 

Fig. \ref{fig:2c} and Fig. \ref{fig:2d} show the calculation for a 11.2 eV photon energy with an active and a transparent surface barrier, respectively. The position of the Fermi level in the calculation has been corrected to match with the experiment and the $k_\perp$ sampled at 21.2 eV (11.2 eV) has been estimated to be approximately 80\% (50\% respectively) of the total $\Gamma Z$ distance. The transparent surface barrier suppresses the surface states and allows us to discriminate the origin of the bands (see Supporting Information for more details about this procedure). The resulting comparison between theory and experiment confirms our attribution of the bands $S_{1,2}$ and $B_1$ to surface and bulk states, respectively. 

Having clarified the nature of the low-energy band structure, we focus now on the bulk state $B_1$ that shows a large energy splitting at low temperature in the ferroelectric phase. We concentrate ourselves now on the data as measured at 11.2 eV photon energy (see Fig. \ref{fig:2b}), which allow to disentangle the bulk states from the surface states and to clearly resolve the bulk Rashba splitting.
By presenting ARPES measurements as a function of temperature, we address the effect of the ferroelectric to paraelectric transition on the amplitude of the bulk Rashba splitting, which is directly correlated to the ferroelectric distortion. 

\begin{figure}[H]
    \includegraphics[width=\linewidth]{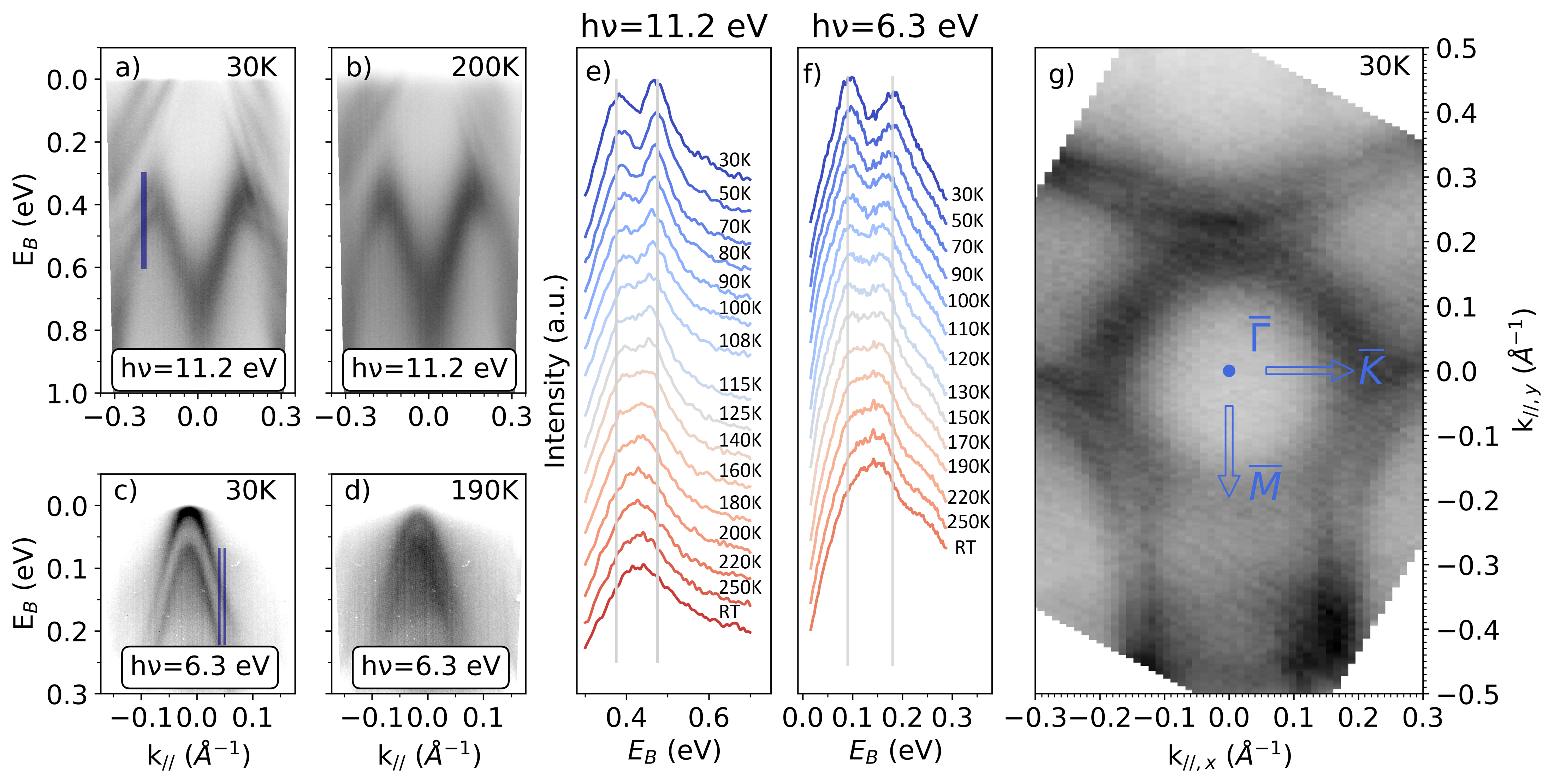}
    \begin{subfigure}[b]{0\linewidth}\phantomcaption\label{fig:3a}\end{subfigure}    \begin{subfigure}[b]{0\linewidth}\phantomcaption\label{fig:3b}\end{subfigure}
    \begin{subfigure}[b]{0\linewidth}\phantomcaption\label{fig:3c}\end{subfigure}
    \begin{subfigure}[b]{0\linewidth}\phantomcaption\label{fig:3d}\end{subfigure}
    \begin{subfigure}[b]{0\linewidth}\phantomcaption\label{fig:3e}\end{subfigure}
    \begin{subfigure}[b]{0\linewidth}\phantomcaption\label{fig:3f}\end{subfigure}
    \begin{subfigure}[b]{0\linewidth}\phantomcaption\label{fig:3g}\end{subfigure}
   \caption{ARPES measurements along the $\overline{K\Gamma K}$ high-symmetry line ($k_{\parallel,y}=0$ \AA$^{-1}$) with a photon energy of 11.2 eV at (a) 30 K and (b) 200 K. ARPES measurements along the line $\overline{K\Gamma K}$ shifted in direction to $\overline{M}$ ($k_{\parallel,y}=-0.13$ \AA$^{-1}$) with a photon energy of 6.3 eV at (c) 30 K and (d) 190 K. (e) EDCs as a function of temperature for a photon energy of 11.2 eV, with k$_{\parallel} \in[-0.20,-0.19]$ \AA$^{-1}$ (blue region in panel (a)). Light-gray lines are guides to the eye for the reader. (f) EDCs as a function of temperature for a photon energy of 6.3 eV, with k$_{\parallel} \in[0.04,0.05]$ \AA$^{-1}$. (g) Constant energy map at binding energy $E_B=0.4$ eV, photon energy $h\nu=11.2$ eV and at $T=30$ K.} 
    \label{fig:3}
\end{figure}

Figures \ref{fig:3a} and \ref{fig:3b} show ARPES spectra taken at 30 K and at 200 K, respectively, with $h\nu=11.2$ eV. At 30 K, a clear splitting is observed in all bands, namely the surface ones $S_{1,2}$ and the bulk one $B_1$. At 200 K, the bands become significantly broader due to thermal effects, but a splitting of the surface-related bands $S_{1,2}$ is still obvious, in contrast to the bulk band $B_1$, for which the splitting is no longer clearly resolved. 
We therefore plot in Fig. \ref{fig:3e} energy distribution curves (EDCs) integrated over k$_{\parallel} \in[-0.20,-0.19]$ \AA$^{-1}$ (blue region in panel Fig. \ref{fig:3a}) for a large temperature range up to room temperature. These EDCs allow us to clearly distinguish the two peaks related to the split bulk band at low temperature (in the ferroelectric phase) and to follow the splitting up to about 160 K. At higher temperature, the two peaks seem to merge together, so that it is difficult to directly assess whether the band splitting persists at high temperatures or whether it collapses.

To answer this question, we have acquired ARPES data using a photon energy of 6.3 eV, to take advantage of the higher momentum resolution at lower photon energy. For this purpose and to maximize the effect of the splitting, we oriented the analyzer slit in a plane parallel to $\overline{K\Gamma K}$, but shifted towards $\overline{M}$ at $k_{\parallel,y}=-0.13$ \AA$^{-1}$ (which allows to see the same bulk band $B_1$ at a different position in the reciprocal space - see Fig. \ref{fig:3g}). 

The corresponding photoemission intensity maps are shown in Fig. \ref{fig:3c} and \ref{fig:3d} for temperatures of 30 K and 190 K, respectively. We observe two hole-like bands with a clear splitting at low temperature (Fig. \ref{fig:3c}). One-step model photoemission calculations confirm that these are the same bulk band $B_1$ (see the calculations with and without a transparent surface barrier in the Supporting Information with Fig. \ref{fig:Appendix}). Moreover, the band splitting is still visible at 190 K (Fig. \ref{fig:3d}).
We have extracted EDCs in this configuration for $k_{\parallel}\in [0.04,0.05]$ \AA$^{-1}$ to follow the reduction of the splitting as a function of temperature (see Fig. \ref{fig:3f}), which allows us to track the splitting up to 250 K at least. 

\begin{figure}
  \begin{subfigure}[b]{0.99\columnwidth}
  \begin{center}
      \includegraphics[height=250pt]{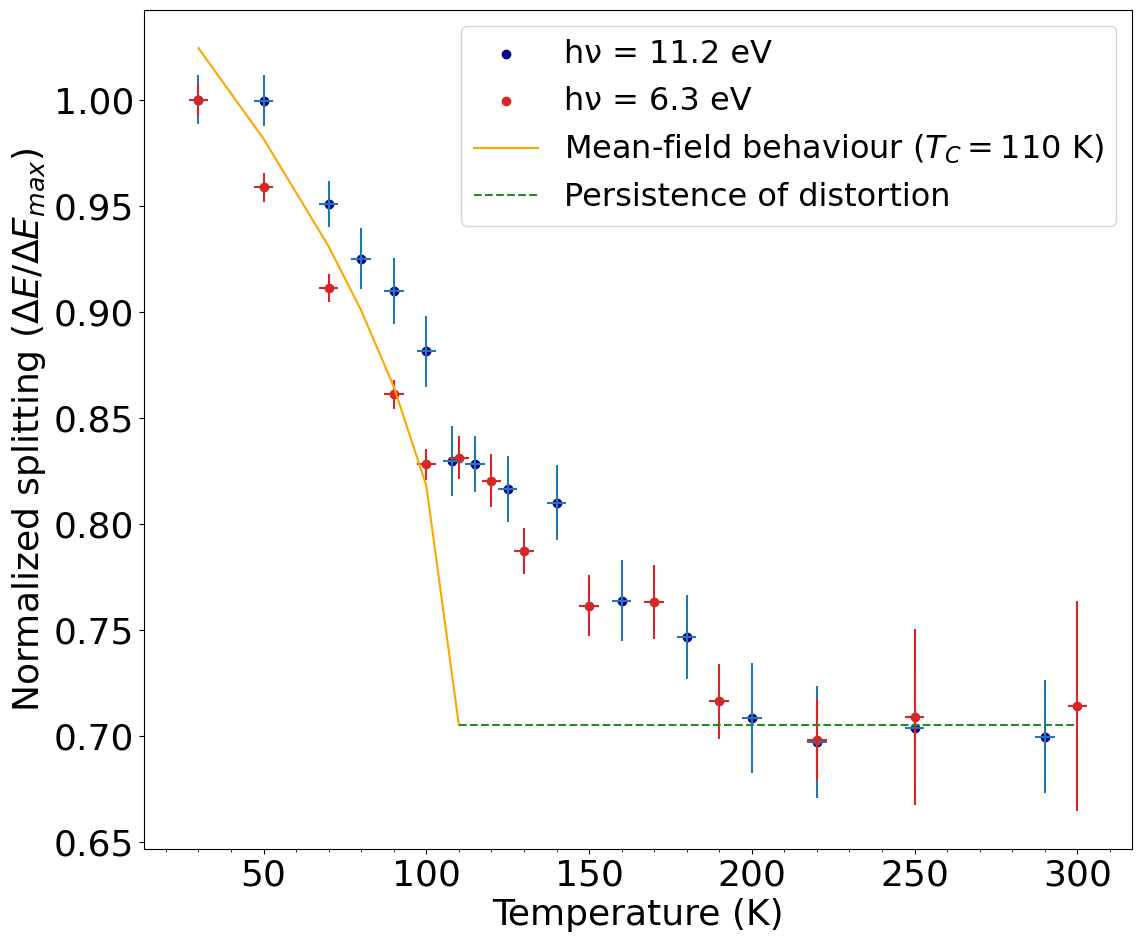}
    \end{center}
   \end{subfigure}
\caption{Evolution of the Rashba bulk band splitting as a function of temperature extracted from ARPES data obtained using a photon energy of 11.2 eV (6.3 eV) in blue (red). A mean-field-like second-order phase transition with a $T_c=110$ K is added on top (orange curve). However, the non-zero splitting above $T_c$ (dashed green line) indicates the persistence of a structural distortion up to room temperature, in disagreement with a mean-field transition. The error bars are due to temperature measurement precision and the fitting procedure.} 
\label{fig:4}
\end{figure}

For a quantitative characterization of the temperature evolution of the bulk states across the ferroelectric transition, we have fitted the EDCs of Fig. \ref{fig:3e} and Fig. \ref{fig:3f} with two Voigt functions (see Supporting Information for more details on the procedure with Fig. \ref{fig:fits}). The variation of the splitting as a function of temperature is plotted in Fig. \ref{fig:4}. We caution that, although we obtained good fits with two Voigt functions for temperatures above 250 K, equally good fits could be obtained with a single broader Voigt function in this temperature range. However, this would lead to an abrupt and nonphysical behavior of the width of the Voigt function around 200 K: therefore we focus on the scenario with two peaks up to room temperature \footnote{A single peak scenario starting above the transition temperature where two peaks are still clearly visible in the EDCs, e.g. at 190 K, gives an absurdly large width. Moreover, this width decreases with increasing temperature, indicating that we are trying to fit with one peak two contributions that are moving closer together.}. 
First of all, we see that the evolution of the band splitting in temperature is the same for the two sets of EDCs, confirming the same mechanism observed with both photon energies. 
Secondly, the reduction of the splitting is particularly pronounced below $100$ K, but a finite value remains at higher temperatures, up to room temperature at odds with the expectation for a paraelectric cubic state at high temperature.
The evolution of the band splitting was reversible and reproducible across different heating and cooling cycles.

\section{\label{sec:Discussion}Discussion}

Although the origin of the transition (displacive vs. order-disorder) remains a matter of debate in the literature, its second-order character is agreed upon \cite{iizumi_phase_1975,pawley_diatomic_1966,brillson_raman_1974,oneill_inelastic_2017,rabe_ab_1985,salje_tin_2010,littlewood_crystal_1980,fornasini_local_2021,knox_local_2014}.
We therefore superimpose on the experimental data in Fig. \ref{fig:4} a mean-field-like (orange) curve with a critical temperature of $T_c=110$ K and adding a constant offset of $\Delta E/E_{max}=0.7$. We stress that, within a mean-field-like scenario, one would expect a zero offset at room temperature. The obtained curve agrees well with the low temperature data, but it reveals a rounding of the phase transition above $T_c$.
 This could be due to strong thermal fluctuations, in agreement with the general increased broadening of the bands observed at and above about 200 K in ARPES (see Fig. \ref{fig:3b}). 
 However, this fails to explain the persistence of a splitting well above $T_c$, which is a clear indication of inversion symmetry breaking inside the crystal even at high temperature.

This surprising observation is consistent with an order-disorder type of phase transition instead of a displacive transition. Whereas for a displacive transition, the onset of the anion/cation displacement appears \textit{only} at $T\leq T_c$ and continuously grows as the temperature decreases, the order-disorder phase transition is based on the freezing of the fluctuations between two equivalent off-center lattice sites at the critical temperature $T_c$. As a result, although above $T_c$ the net macroscopic polarization is zero, there are still clusters with a non-zero local polarization extending over a few-unit cells and with alternation of the sign of the polarization from cluster to cluster. These clusters would therefore still have locally a structural distortion and thus give rise to the spin-split bulk bands above $T_c$ as we experimentally observe. This agrees with extended x-ray absorption fine structure, x-ray scattering  analysis of the pair distribution function that have revealed the persistence of local lattice distortions, i.e. the presence of local ferroelectric dipoles above $T_c$ \cite{fornasini_local_2021,mitrofanov_local_2014,fons_phase_2010,matsunaga_order-disorder_2011}.
We caution though that we cannot rule out another structural mechanism occurring specifically near the surface of SnTe that could cause the persistence of the band splitting at high temperature\cite{aggarwal_local_2016}, given that ferroelectricity has been observed up to room temperature in the two dimensional limit of SnTe \cite{chang_discovery_2016} .

From our data at a photon energy of 11.2 eV (see Fig. \ref{fig:2a}), we can estimate a value for the Rashba parameter $\alpha_R$. With the standard relation $\alpha_R=2 E_R / k_0$, we extract $E_R=0.34$ eV and $k_0=0.19$ \AA$^{-1}$ at 30 K. The Rashba parameter is then $\alpha_R=3.58 $ eV \AA. 
We note that this experimental value is relatively close to the theoretical estimation from DFT in Ref. \cite{plekhanov_engineering_2014} ($\alpha_R=4.4 $ eV \AA), providing therefore an experimental confirmation of the giant Rashba effect in SnTe. 

Given our discovery of the persistence of a structural distortion in SnTe at higher temperatures, an open question is what is its impact on the topological surface states? 
Symmetry arguments have been used to derive a non-zero mirror Chern number on the (001), (111) and (110) surfaces of the rocksalt paraelectric structure, and therefore the presence of Dirac cones in the ARPES spectra \cite{fu_topological_2011}. Such considerations were supported by earlier theoretical \cite{hsieh_topological_2012,yan_experimental_2014} and experimental studies on the (001) surface \cite{tanaka_experimental_2012}. As for the (111) orientation, static ARPES studies have also claimed to have measured a topological surface state at $\overline{\Gamma}$ \cite{tanaka_two_2013,zhang_arpes_2017,yan_experimental_2014}. However, the highly p-type character of SnTe precludes the direct observation of the Dirac cone by ARPES. Our results provide a new perspective to these findings by resolving more bands, namely two surfaces states instead of one, and with an unprecedented resolution. By looking at our ARPES measurements, we identify the $S_1$ and $S_2$ surface states as the candidate for the linear dispersion in the occupied states that has been interpreted as a topological surface state in previous studies. Furthermore, our one-step calculation (see e.g. Fig. \ref{fig:2c}) confirm the finding of Plekhanov \textit{et al.} \cite{plekhanov_engineering_2014}, which shows that in the rhombohedral structure, there is no topological surface state near the Fermi level. Based on our observation of the persistence of local lattice distortions above $T_c$, we therefore expect that no topological surface states appear at high temperature on the (111) surface, at odds with recent time-resolved ARPES studies \cite{ito_observation_2020}. Our new results therefore requires a reassessment of these observations.

\section{\label{sec:Conclusion}Conclusion}
We have characterized the band structure of SnTe(111) using high-energy resolution ARPES measurements with an unprecedented quality. Combined with state-of-the-art photoemission calculation with and without a surface barrier, our ARPES study at selected photon energies enabled us to differentiate surface and bulk states. The presence of bulk-split bands has been directly connected to inversion symmetry breaking. We have also studied the evolution of this splitting as a function of temperature to characterize the ferroelectric transition. This study demonstrated inconsistencies with a displacive mean-field like transition, revealing a rounding of the phase transition and a splitting persisting above $T_c$, at least up to 250 K.
This observation is consistent with an order-disorder type phase transition, in agreement with findings from other studies using local probes \cite{fornasini_local_2021,mitrofanov_local_2014, aggarwal_local_2016}. Above the critical temperature, fluctuations of the polarization vector from one cluster to another implies that a structural distortion remains and explain the persistence of the band splitting at high temperature. 
We propose that the possible persistence of ferroelectricity at high temperature in the near-surface region could be tested with spin-resolved and microfocus ARPES measurements by looking for the existence of a finite spin polarisation, as well as by evidencing circular dichroism in ARPES \cite{PhysRevLett.121.186401,kim_prediction_2019}. 
Finally, the persistence of rhombohedral distortions above the critical temperature requires a reassessment of the topological nature of the SnTe(111) surface, since it has been shown in the literature \cite{plekhanov_engineering_2014} and confirmed by our DFT calculation that the break of symmetry destroys the topological surface state along the (111) direction.

\begin{acknowledgement}
Aki Pulkkinen and Geoffroy Kremer contributed equally to this work.

G.S. would like to thank the Austrian Science Fund (FWF), who supported this study with projects No. P30960-N27 and I 4493-N. 
We are greatful to Natalia Olszowska and Jacek Kołodziej for their support of the ARPES measurements at SOLARIS, funded by the Polish Ministry of Education and Science under contract nr $1/SOL/2021/2$.
J. M. and A. P. would like to thank the CEDAMNF project with reg. no. $CZ.02.1.01/0.0/0.0/15_003/0000358$ and the QM4ST project with reg. no. $CZ.02.01.01/00/22_008/0004572$, co-funded by the ERDF as part of the MŠMT.

We are very grateful to M. Rumo and B. Salzmann for fruitful discussions. Skillful technical assistance was provided by F. Bourqui, B. Hediger and M. Audrey.
\end{acknowledgement}

\section{\label{sec:Methods}Methods}

\subsection{Sample growth}
Epitaxial SnTe(111) films of 2 µm thickness were grown by molecular beam epitaxy on BaF$_2$ substrates under ultra-high vacuum (UHV) conditions at a substrate temperature of 350$^{\circ}$ C and a compound effusion cell. During growth, the SnTe(111) surface exhibits a perfect two-dimensional reflection high-energy electron diffraction pattern revealing a perfect 2D growth mode. After growth, the samples were transferred to the ARPES setup without breaking UHV conditions using a battery operated vacuum suitcase having a pressure of better than $10^{-10}$ mbar. It is noted, that due to the high density of native Sn vacancies, SnTe intrinsically exhibits a high p-type carrier concentration of typically  $2$ x $10^{20}$ cm$^{-3}$. 
For this reason the Fermi level is always inside the topmost valence band. The lattice parameter of the SnTe layers was determined to be $a = 6.325$ \AA at room temperature (rhombohedral lattice parameter of $4.472$ \AA) , which is in good agreement with literature values\cite{iizumi_phase_1975}.

\subsection{ARPES measurements}
Temperature-dependent angle-resolved photoemission spectroscopy (ARPES) investigations were carried out using a Scienta DA30 photoelectron analyzer with a base pressure better than $3 \times 10^{-11}$ mbar. Photons sources are monochromatized He$_I$ (and Xe) radiation with $h\nu = 21.22$ eV ($h\nu = 11.2$ eV) and a high energy-resolution laser based on a commercial setup (Harmonix, APE GmbH) generating 6.3 eV photons using
harmonic generation from the output of an optical parametric oscillator pumped by a Paladin laser (Coherent,
inc.) at 80 MHz. The total energy resolution was about 10 meV and cooling of the sample was carried out at rates $<$2 K/min to avoid thermal stress. Each measurement was preceded by a break of at least 15 min, to ensure thermalisation. Accordingly, the error on the absolute sample temperature was estimated to be well below 5 K.
The photon-energy dependent ARPES measurements were performed at $30$ K at the URANOS beamline of the SOLARIS synchrotron in Krakow using a Scienta DA30L photoelectron analyzer.

\section{Supporting Information}
\subsection{Photoemission calculations}
The ARPES calculations were performed using the one-step model of photoemission implemented in the multiple scattering Green's function code SPRKKR \cite{Ebert_2011,BRAUN20181}. The bulk electronic structure was calculated within the atomic sphere approximation with angular momentum expansion up to $l_{\rm max} = 3$ with lattice parameter $a=4.4547$ \AA, rhombohedral distortion angle 59.9 degrees, and $z_{\rm Te} = 0.52$. To investigate the origin of the spectral features, we have compared one-step model calculations with the Rundgren-Malmström model surface barrier and a transparent barrier, which allows us to evaluate contributions from surface-related bands \cite{JRundgren_1977}. 

In the layer-KKR formalism of the one-step model of photoemission\cite{pendry1976,braun2018}, the crystal structure is divided into layers whose transmission and reflection factors characterize the photocurrent attenuation inside the crystal. In addition to the atom layers, the Rundgren-Malmstr\"om\cite{malmstrom1980} type surface barrier connecting the inner potential of the crystal to the vacuum level is also treated as a layer with transmission and reflection factors. The surface barrier layer reflection factor is set to zero in the transparent barrier calculations.

\subsection{Density functional theory calculations}
Density functional theory (DFT) band structure calculations with spin-orbit coupling were performed using the Vienna ab-initio software package (VASP) \cite{PhysRevB.47.558,PhysRevB.49.14251,KRESSE199615,PhysRevB.54.11169,PhysRevB.59.1758}. The SnTe(111) surface was modeled as a repeated slab geometry of 72 atom layers and 24 \AA\, of vacuum between the adjacent slabs. Exchange and correlation (xc) effects were treated at the level of the generalized gradient approximation using the PBE xc-functional \cite{PhysRevLett.77.3865}. The kinetic energy cutoff was set to 250 eV, and a 9×9×1 $k$-point mesh was used for the Brillouin zone sampling in the slab calculations. 

\subsection{\label{sec:Appendix_hnuARPES} Photon-energy dependent ARPES measurements}

\begin{figure}[H]
      \begin{center}
      \includegraphics[width=0.9\columnwidth]{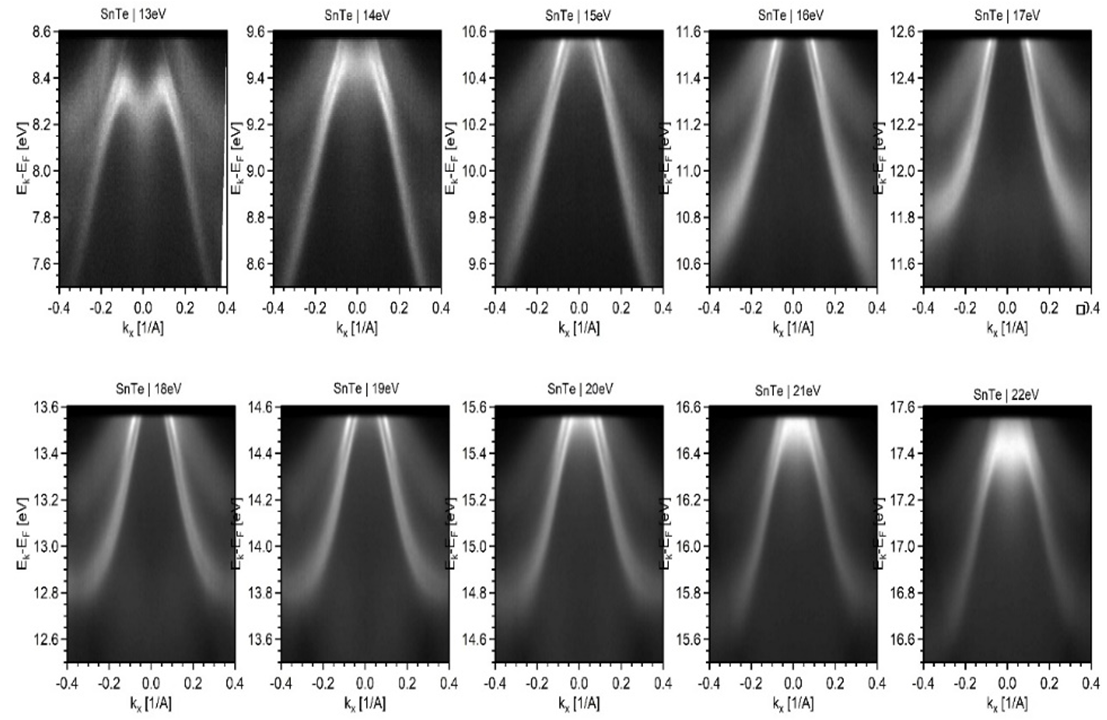}
      \end{center}
\caption{Photon-energy dependent ARPES measurements along the $\overline{K\Gamma K}$ high-symmetry direction at $30$ K with a photon energy ranging from 13 to 22 eV as indicated.} 
\label{fig:hvdep}
\end{figure}

Figure \ref{fig:hvdep} shows photon-energy dependent ARPES measurements taken along the $\overline{K\Gamma K}$ high-symmetry direction at 30 K with photon energy ranging from 13 to 22 eV. These data allows to follow the $k_\perp$ dispersion of the different bands identified in the main text. Although the intensity of the surface states $S_1$ is lower than in the main text (probably due to a slightly degraded surface quality), we observe that the states $S_1$ and $S_2$ do not disperse as function of photon energy, contrarily the state $B_1$. These data confirms our identification of $S_1$ and $S_2$ as surface states and $B_1$ as a bulk state.

\subsection{\label{sec:Appendix_6eV} Photoemission calculation for the 6.3 eV ARPES data}

\begin{figure}[H]
  \begin{subfigure}[b]{0.99\columnwidth}
      \phantomcaption\label{fig:5a}
      \begin{center}
      \includegraphics[width=0.7\columnwidth]      {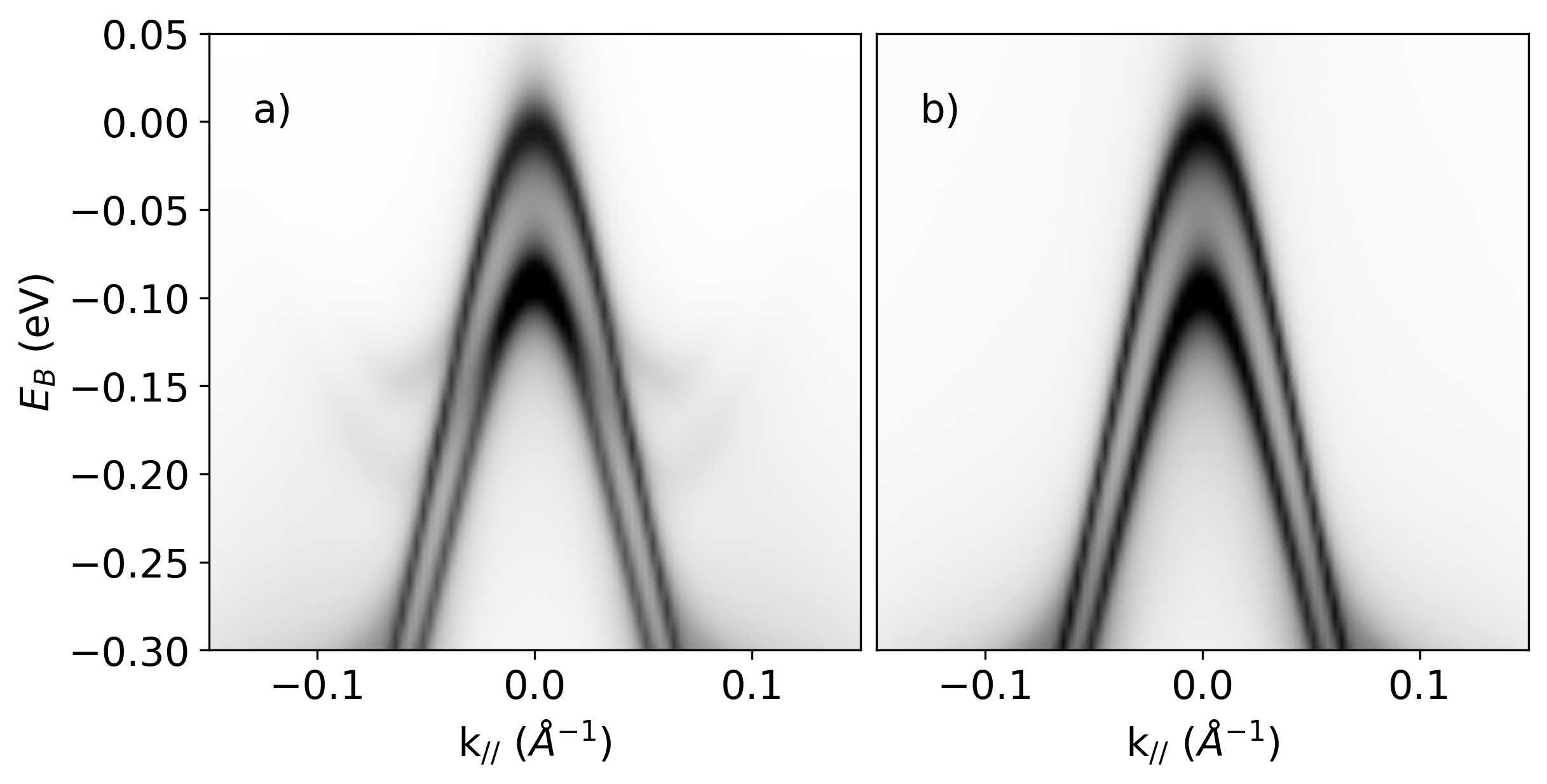}
      \end{center}
      \phantomcaption\label{fig:5b}
   \end{subfigure}
\caption{One-step photoemission calculations for a slab geometry in a ferroelectric state, with a photon energy of 6.3 eV and a Te termination, along a line parallel to $\overline{K\Gamma K}$ shifted towards $\overline{M}$ at $k_{\parallel,y}=-0.13$ \AA$^{-1}$ without (a) and with (b) a transparent surface barrier.} 
\label{fig:Appendix}
\end{figure}

Figure \ref{fig:5a}  and \ref{fig:5b} displays the one-step photoemission calculation with and without, respectively, the surface barrier at a photon energy of 6.3 eV along the same wave vector direction as used in the experiment. Despite slight differences with the experimental data in Fig. 3c of the main text (e.g., a very weak surface state in the calculations that seems to be hidden in the background of the measurement), the qualitative comparison is good and proves that the investigated band is a bulk one.

\subsection{\label{sec:AppendixsurfTerm} DFT calculations for different surface terminations}

\begin{figure}[H]
  \begin{subfigure}[b]{0.35\linewidth}
      \caption{Sn terminated}\label{fig:6a}
      \includegraphics[width=\columnwidth]{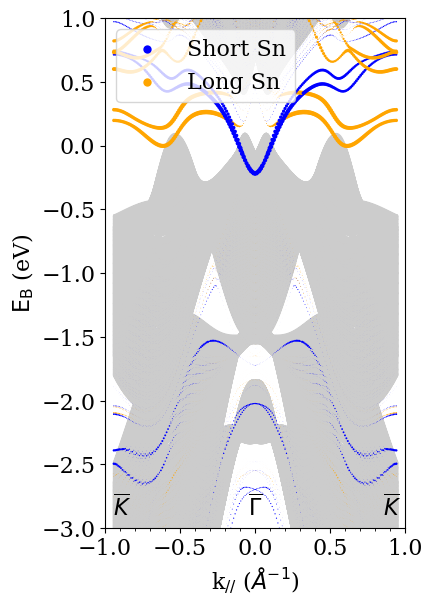}
   \end{subfigure}
   \begin{subfigure}[b]{0.35\columnwidth}
        \caption{Te terminated}\label{fig:6b}
        \includegraphics[width=\columnwidth]{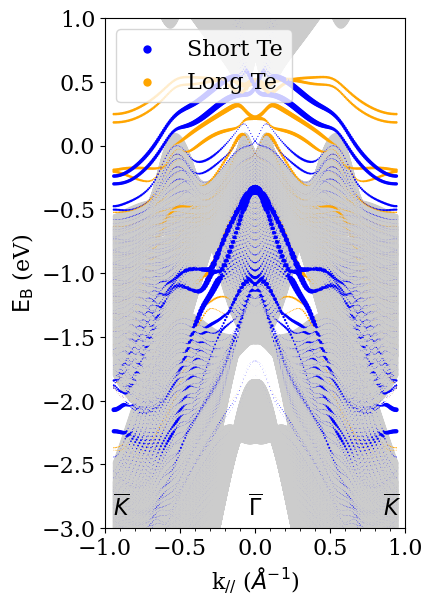}
    \end{subfigure}
\caption{(a) DFT calculation of SnTe(111) made with VASP with a (a) Sn surface termination and (b) a Te surface termination. The shaded area is the bulk projected bands and the size of the blue and yellow markers represents the contribution of the top surface atoms to the character of the bands for the long and short configurations, respectively (see text).}
\label{fig:6}
\end{figure}

Figure \ref{fig:6} displays the DFT calculation of SnTe(111) in its ferroelectric state for four different terminations, namely the Te or Sn terminated surfaces, and for each case with a short or a long distance between the first two atomic planes (called short and long configurations thereafter). The bulk projected bands are integrated in the grey area and the contribution of the first surface layer is highlighted in yellow or blue dots, for the long and short configurations, respectively. As such, we expect to visualize the surface state $S_1$ (by analogy with GeTe, $S_2$ being interpreted as a surface resonance state having a deeper origin in the crystal \cite{krempasky_disentangling_2016}). The comparison with the experimental data (see Fig. 2 of the main text) confirms our interpretation of the origin of the bands and clearly proves that our samples are Te terminated. The short distance configuration seems more likely, because of the presence of surface states with higher binding energies and because its work function (5.05 eV versus 5.45 eV for the long configuration) is closer to the experimental value (4.65 eV).

\section{\label{sec:Appendix_rashba} Fitting of the Rashba splitting}

\begin{figure}[H]
      \begin{center}
      \includegraphics[width=0.99\columnwidth]{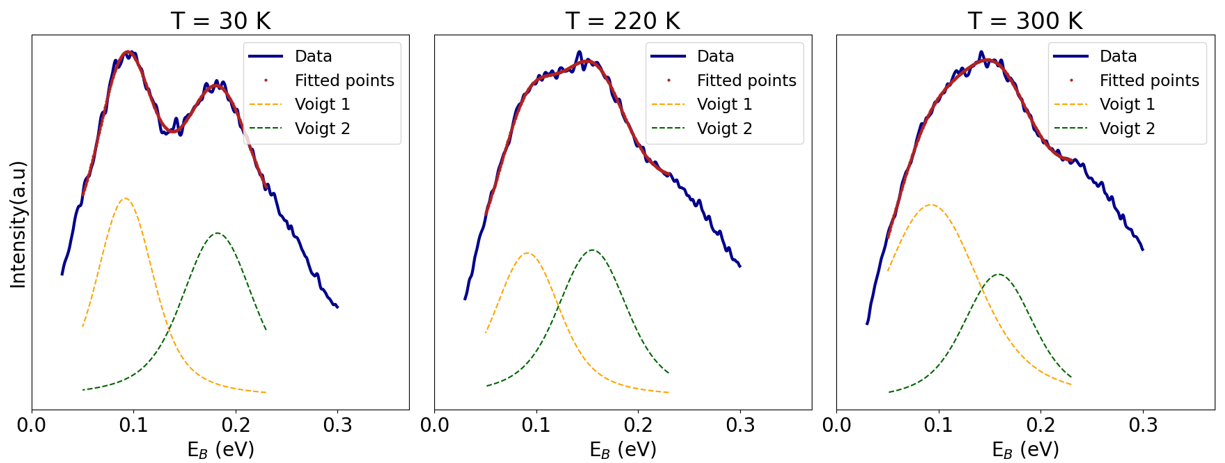}
      \end{center}
\caption{Exemplary fits (red) the EDCs (blue) at a photon energy of 6.3 eV to extract the Rashba splitting as a function of temperature. The orange and the green curves represents the two contributions of a Voigt function. } 
\label{fig:fits}
\end{figure}

Figure \ref{fig:fits} illustrates how the fits of the energy distributions curves (EDCs) were done in order to extract the Rashba splitting as a function of temperature. The combination of two Voigt functions with a linear background was used to fit the EDCs on a limited energy range, using the Levenberg-Marquardt algorithm of the lmfit software \cite{newville_matthew_2014_11813}. No other constraints were imposed on the fit, apart from a manually chosen first guess. The splitting is determined as the distance between the centers of the two Voigt functions.

\bibliography{MyLibrary}

\end{document}